\documentclass[npg]{copernicus}
\usepackage{graphics,graphicx,dcolumn,bm,fleqn,epic,eepic,epsfig}
\usepackage{amssymb,amsmath,multirow,rotate,color,float}
\usepackage{times}
\usepackage{color}
\usepackage{soul}                    
\DeclareMathOperator{\sgn}{sgn} 
\definecolor{red}{rgb}{1,0,0}
\definecolor{green}{rgb}{0,1,0}
\definecolor{blue}{rgb}{0,0,1}

\begin{document}

\title{Size distribution and structure of Barchan dune fields}

\author[1]{Orencio Dur\'an}
\author[2]{Veit Schw\"ammle}
\author[3]{Pedro G.~Lind}
\author[4,5]{Hans J.~Herrmann}

\affil[1]{Laboratoire de Physique et M\'ecanique des Milieux 
          H\'et\'erog\`enes, (CNRS UMR 7636), 
          ESPCI, 10 rue Vauquelin, 75231 Paris Cedex, France}
\affil[2]{Department of Biochemistry and Molecular Biology,
          University of Southern Denmark, 
          Campusvej 55, DK-5230  Odense M, Denmark}
\affil[3]{Centro de F\'{\i}sica Te\'orica e Computacional, 
          Faculdade de Ci\^encias da Universidade de Lisboa,
          Av.~Prof.~Gama Pinto 2, 1649-003 Lisboa, Portugal}
\affil[4]{Computational Physics, IfB, HIF E12, ETH H\"onggerberg,
          Schafmattstr.~6, CH-8093 Z\"urich, Switzerland}
\affil[5]{Departamento de F\'{\i}sica, Universidade Federal do Cear\'a,
          60451-970 Fortaleza, Cear\'a, Brazil}


\runningtitle{Structure of dune fields}

\runningauthor{Duran et al}

\correspondence{Pedro G.~Lind\\ (plind@cii.fc.ul.pt)}

\received{}
\pubdiscuss{} 
\revised{}
\accepted{}
\published{}


\firstpage{1}

\maketitle

\begin{abstract}
Barchans are isolated mobile dunes often organized in large dune fields. 
Dune fields seem to present a characteristic dune size and spacing, 
which suggests a cooperative behavior based on dune interaction. 
In Duran {\it et al.} (2009), we propose that the redistribution of sand 
by collisions between dunes is a key element for the stability and size 
selection of barchan dune fields. 
This approach was based on a mean-field model ignoring the spatial 
distribution of dune fields.
Here, we present a simplified dune field model that includes the 
spatial evolution of individual dunes as well as their interaction through 
sand exchange and binary collisions. 
As a result, the dune field
evolves towards a steady state that 
depends on the boundary conditions. 
Comparing our results with measurements of Moroccan dune fields, we find 
that the simulated fields have the same dune size distribution as in real
fields but fail to reproduce their homogeneity along the wind direction.
\end{abstract}



\introduction

\begin{figure*}[htb]
\vspace*{2mm}
\begin{center}
\includegraphics[width=0.8\textwidth]{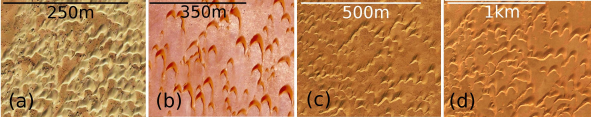}
\includegraphics[width=0.17\textwidth]{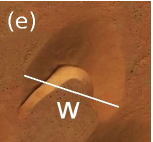}

\includegraphics[width=0.49\textwidth]{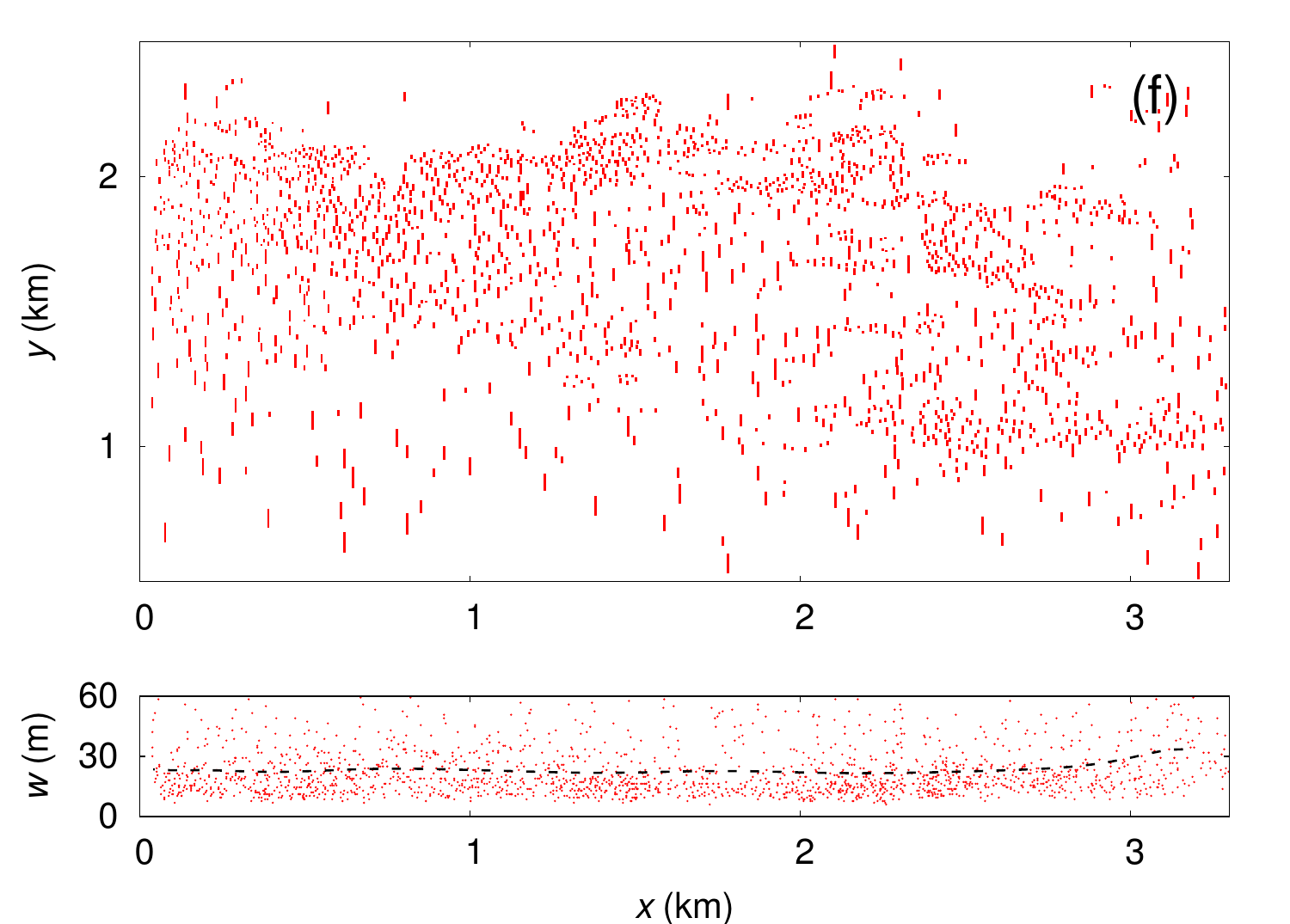}
\includegraphics[width=0.49\textwidth]{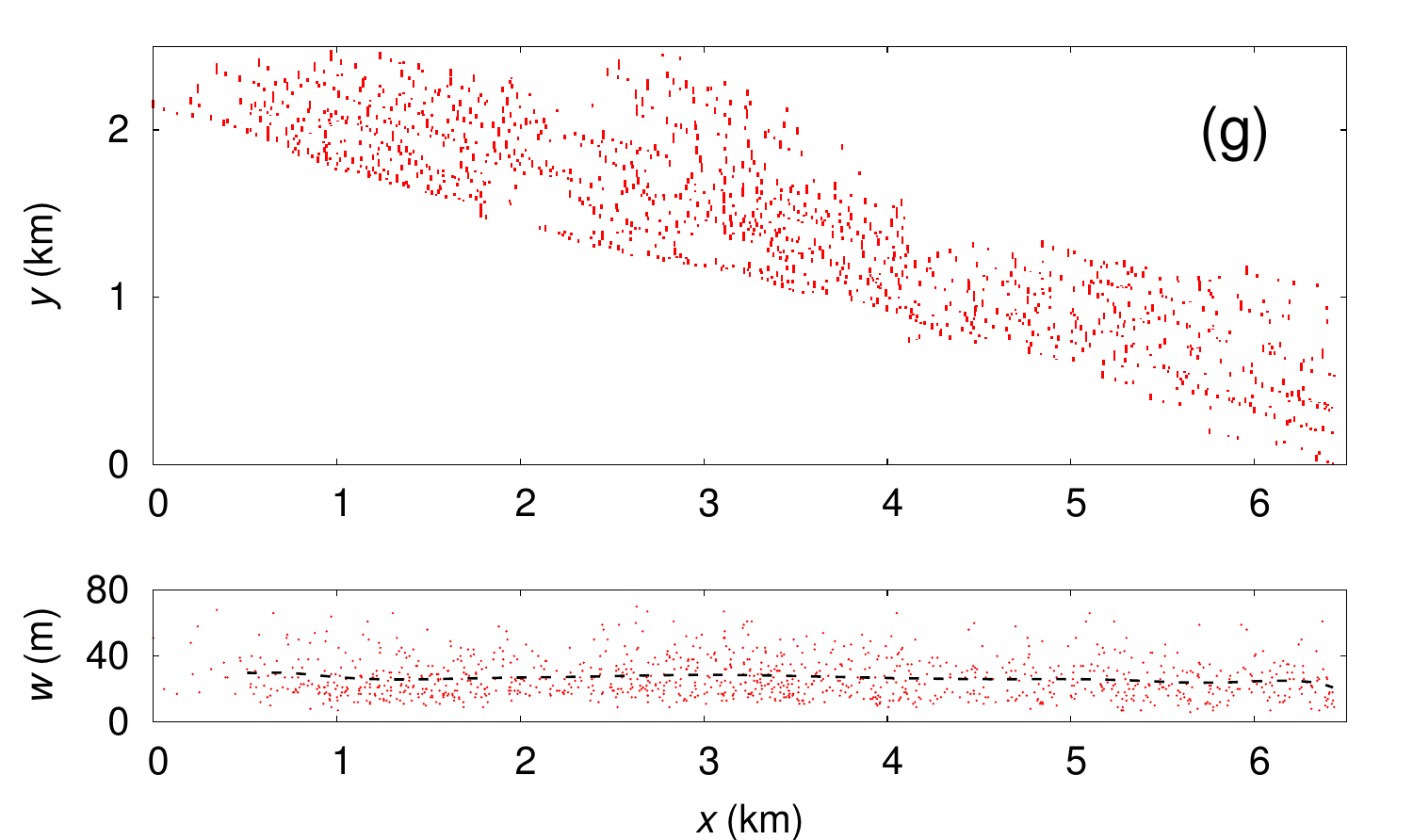}

\includegraphics[width=0.49\textwidth]{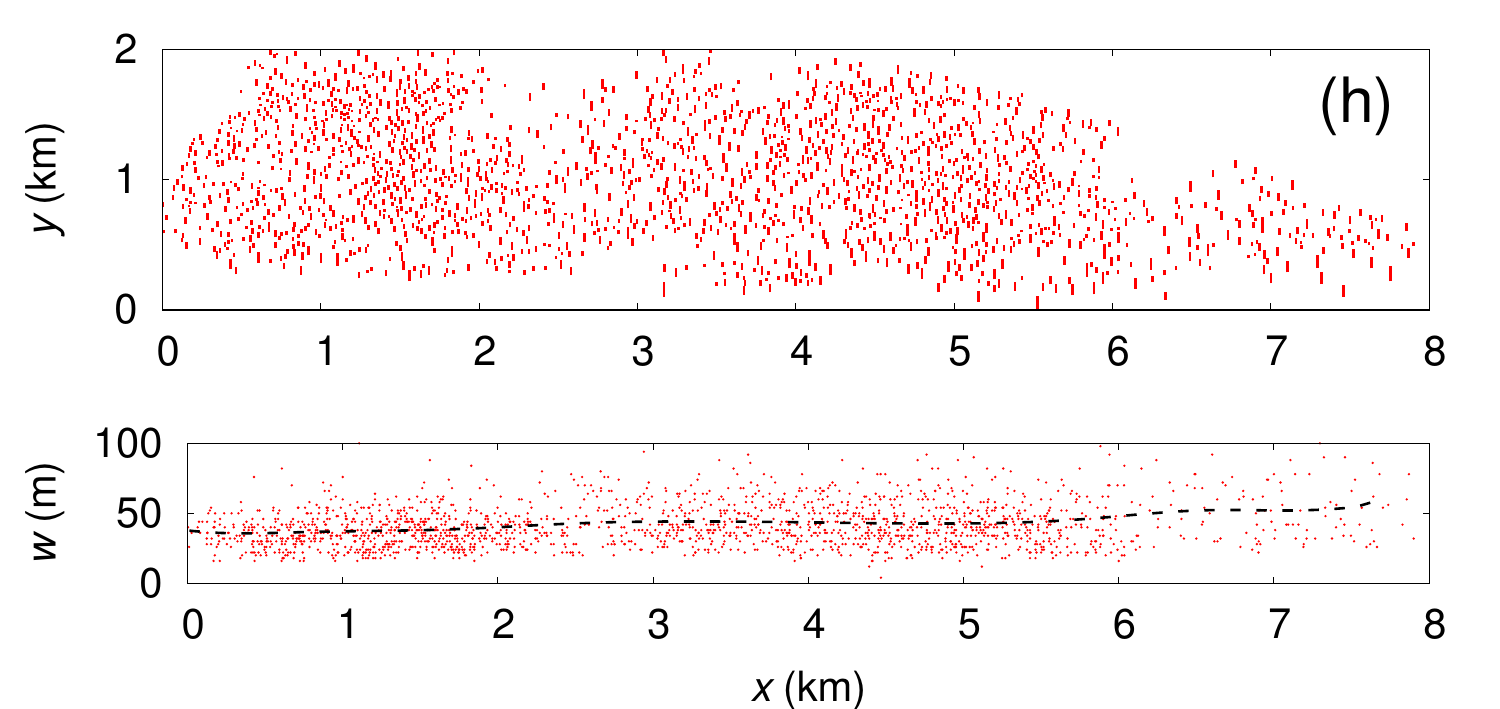}
\includegraphics[width=0.49\textwidth]{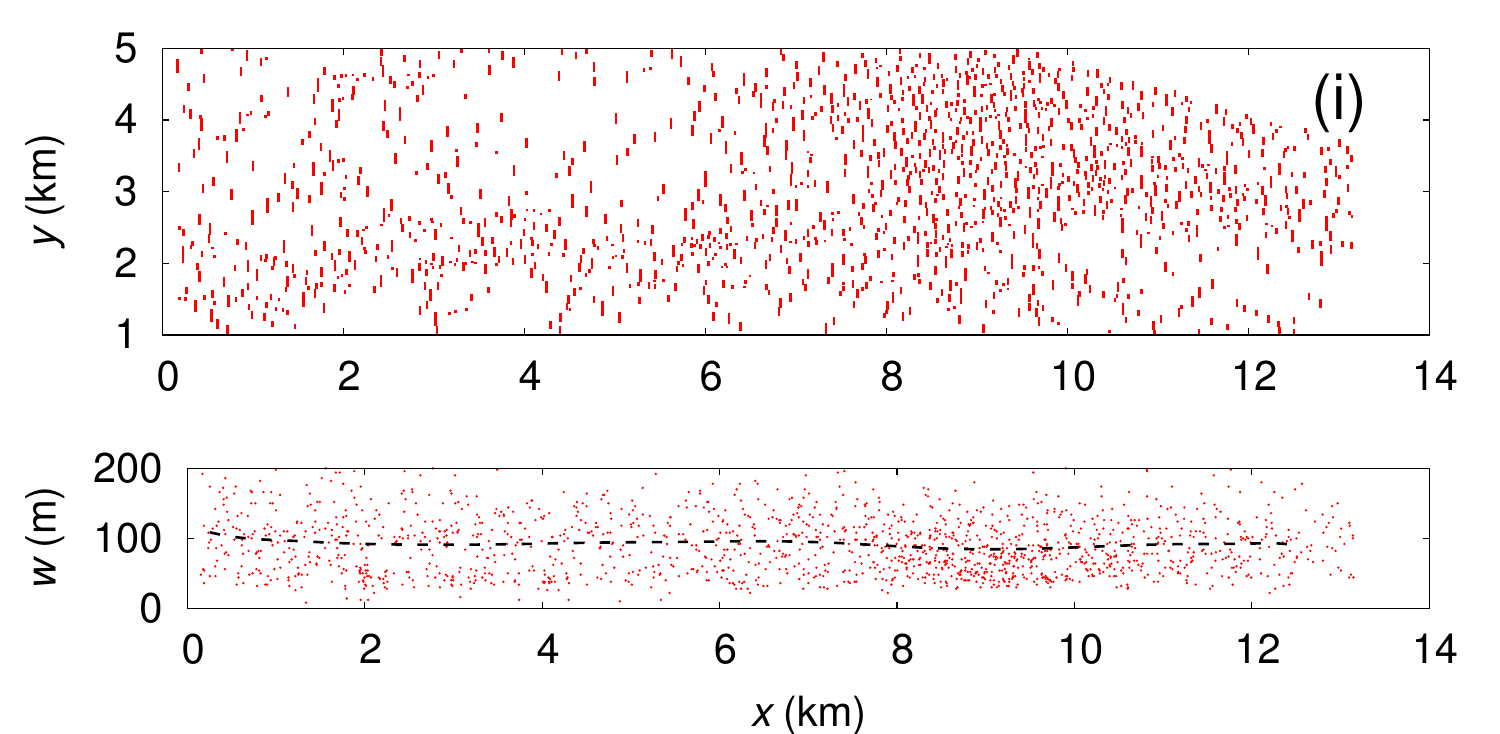}
\end{center}
\caption{(Color online) 
     {\bf (a)-(d)} Detail of the four measured dune field corridors 
           localized in Western Sahara (wind blows from top to bottom). 
     {\bf (e)} Dunes are represented by their `width line'.
     {\bf (f)-(i)} Measured barchan dunes in the four corridors.
          The $x$ axis is oriented along the wind direction 
          (from left to right) which is also the direction of dune movement. 
          Lower panels show the dune width as function of their $x$-position, 
          where dashed lines indicate the average dune size along the 
          transversal $y$-direction.
          In all pictures, the North points up.}
\label{fig:fieldReal}
\end{figure*}

Barchan dunes can be found in fields with low sand availability and 
unidirectional wind. Above their minimum height, of about one meter, 
they show regular shapes with simple scaling relations between their height, 
width, length and 
volume [\cite{AndreottiClaudin02,Elbelrhiti08,Parteli07a}]. Moreover, 
the velocity of one barchan is proportional to the inverse of its
width [\cite{Hersen04}].
Barchan dunes generally do not appear isolated but instead belong to several 
kilometer long dune fields, forming corridors oriented along the wind direction. 
Within these corridors the dunes show rather well selected sizes and 
inter-dune spacing (see Fig.~\ref{fig:fieldReal}(a)-(d)). 
However, single barchans alone are intrinsically unstable 
and they either continuously grow or shrink. 
This discrepancy leads us to the assumption that, at the statistical level, 
the behavior and evolution of single dunes results from the interaction with 
their surroundings typically composed of several thousand 
dunes [\cite{Hersen04b,Elbelrhiti05}]. 

Collisions between dunes have been proposed to be one of the processes 
responsible for the stability of dune 
fields [\cite{SchwaemmleHerrmann03,Duran05,Hersen05b}], another one being dune 
calving due to wind fluctuations [\cite{Elbelrhiti05}]. 
In a recent work, we have already shown that binary collisions {\it alone} 
behave as an additive random process that leads to a stationary Gaussian 
dune size distribution [\cite{Duran09}]. We also found that, after adding sand 
flux exchange into a mean-field model for the evolution of the dune size 
distribution, the collision-based Gaussian distribution transforms into a 
new distribution which is similar in shape to a log-normal one [\cite{Duran09}]. 
In this mean-field approach however, we ignored the spatially extended 
character of mobile dune fields. 
The model, due to its restrictions, does not provide explanation to several 
issues. For instance, it is still not clear which conditions lead to the 
different characteristic sizes in different dune field corridors. Therefore, 
a more realistic approach is used in the work presented here.

In this paper, we present further details on dune collisions
and proceed to model an entire dune field 
based on the scaling relations of isolated dunes. These relations were 
extracted from simulations using a continuous sand flux balance 
model [\cite{SauermannKroy01,KroySauermann02,SchwaemmleHerrmann05}]. 
The aim is to highlight the underlying processes that may lead to size 
selection in a dune field. 
In addition, we carry out a quantitative analysis on how the
external conditions influence the dune field comparing results from
simulation with empirical ones.

The paper is organized as follows:
In Sec.~\ref{sec:measures} we present 
the measured size distributions and inter-dune spacing distributions 
for real fields, namely in four barchan dune fields along 
the coast of Western Sahara (Fig.~\ref{fig:fieldReal}).
We show that spatial homogeneity is an ubiquitous feature of dune fields,
which present a clear characteristic dune size and a well defined inter-dune 
spacing.
In Sec.~\ref{sec:DUNE} we start by describing a simulation of a whole 
dune field using a continuous sand flux balance model
which reproduces qualitatively the real dune field.
Then, observing that the number of dunes is too small 
to be statistically relevant, 
we present in Sec.~\ref{sec:collisions} a model for the internal dynamics of 
the dune field taking into account only binary collisions.
In Sec.~\ref{sec:simple}, this description turns out to be 
oversimplified, motivating the further inclusion of both, the simple rules for barchan 
collisions and the sand flux balance on isolated dunes,
and therefore introducing a simplified model of a large dune field.
Simulations of such model are presented in Sec.~\ref{sec:results}, providing
scaling relations between the spatial distribution, the size distribution and 
the boundary conditions. 
Finally, 
the conclusions are presented,
with additional discussions on dune calving [\cite{Elbelrhiti05}] in the
scope of the stability of dune fields.

\section{Empirical data and data analysis}
\label{sec:measures}

\subsection{Data sets}

The Moroccan desert in Western Sahara contains the longest barchan 
dune fields on Earth.
Satellite images of these deserts are good sources for statistical input 
to calculate the size distribution of sand dunes. 
In Ref.~[\cite{Duran09}] the distribution functions of dune sizes have 
already been presented. 
Here we are particularly interested in the spatial distribution of the 
dunes.

In Western Sahara, barchan dunes develop under a strong uni-directional 
wind in tens of kilometers long corridors with, at least over reasonable 
large regions, a characteristic dune size and a homogeneous dune 
distribution (Fig.~\ref{fig:fieldReal}(a)-(d)). 

It has been shown, both from models~[\cite{SauermannPhD01,Hersen04}]
and measurements~[\cite{Sauermann00,Elbelrhiti05,Elbelrhiti08}], that 
the velocity of barchan dunes as well as height, area and volume, are well 
characterized by their width $w$ solely. 
Therefore, we only measure the width and position of more than $5,000$ dunes 
corresponding to four dune field corridors between Tarfaya and Laayoune 
(Morocco) using satellite images from GoogleEarth, with one meter per pixel 
resolution.  

The four dune fields illustrated in Fig.~\ref{fig:fieldReal} have
respectively $1295$, $1113$, $1947$ and $1630$ barchan dunes, 
covering areas of $\sim 3, 7, 12$ and $60$ km$^2$ and with average 
dune sizes of $17$m, $27$m, $42$m and $86$m  respectively.
The width line is defined as the 
largest distance between the dune horns, as illustrated in 
Fig.~\ref{fig:fieldReal}e. 
Figure~\ref{fig:fieldReal}(f)-(i) 
shows the four measured dune fields, where each 
barchan is represented by its `width line' as a function of its $x$-coordinate 
(downwind distance) along the corridor.
The downwind direction in a barchan dune field is given by the horns
of the dunes[\cite{SauermannKroy01}].

The errors of the measured widths and location of the dunes are of the
same order as the resolution of the satelite image, namely $1$m, which is 
in most cases neglegible in comparison with the width. Therefore we
do not consider such errors here.
In the cases where only one horn is visible, the width is taken assuming
the dune to have a symmetrical shape. Calving is therefore not considered
in our analysis.
Further, a set of overlaping dunes (see Fig.~\ref{fig:RealCollAll})
is either neglected or taken as a single dune in case one dune is
much larger then the other ones.

From Fig.~\ref{fig:fieldReal}, one sees that there is no clear trend in 
the spatial distribution at the scale of the image resolution.
One also notes that, while between corridors a wide variety of dune 
widths is observed, ranging from 5~m to 250~m, together with different 
dune concentration, each corridor {\it per se} has a characteristic dune 
size. 

\subsection{Features of empirical data}

All four measured dune fields have a common underlying size 
distribution function, close to a log-normal and is well reproduced by a 
master equation that balances the dune growth due to sand flux exchange with 
the sand redistribution due to collisions between dunes [\cite{Duran09}]. 

While the size distribution can be fully described only by the mean 
size (see [\cite{Duran09}]), what determines the characteristic size 
at different corridors is still unknown. 

The spatial distribution of dunes provides additional information
beyond the size distribution, namely about the sand distribution within 
the field and the total amount of sand transported through it [\cite{Duran09}].

\begin{figure}[htb]
\vspace*{2mm}
\begin{center}
   \includegraphics[width=0.4\textwidth]{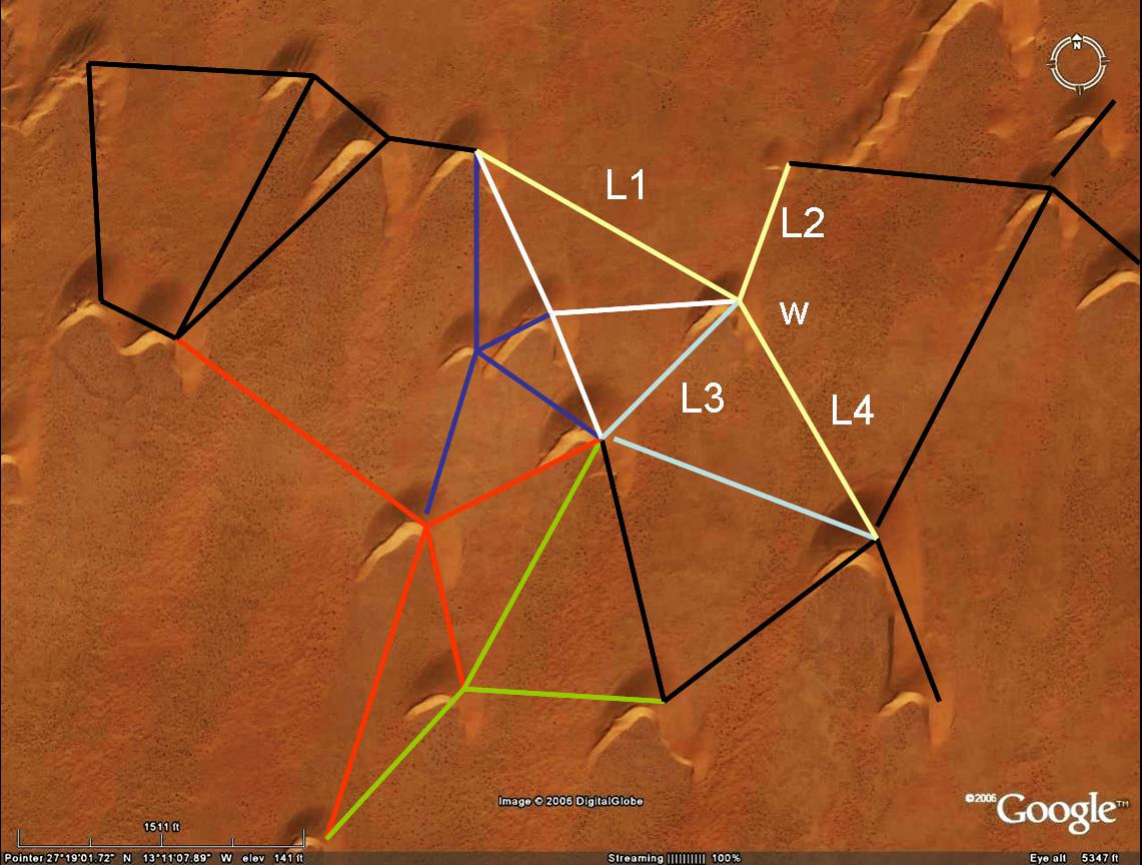}
\end{center}
\caption{(Color online) 
         Dune field planar network. The neighborhood of a dune of width 
         $w$ is defined by its nearest neighbors in each of the four 
         quadrants placed at a distance $L_1$, $L_2$, $L_3$ and $L_4$, 
         respectively.} 
    \label{fig:fieldnetwork}
\end{figure}

We define the inter-dune spacing, $L(w)$, as the characteristic distance between a dune of width $w$ and its neighbors,
\begin{equation}
L(w)\equiv \sqrt{A_f(w)}~,
\end{equation}
where $A_f(w)$ is the sand-free area around the dune. 
This area is computed as follows.
Each dune is connected to its four nearest neighbors, one at each
quadrant of a Cartesian coordinate system centered at the dune,
composing a planar dune network as sketched in Fig.~\ref{fig:fieldnetwork}. 
After searching the nearest neighbors of each dune, the edges joining
the neighbors of one particular dune compose a polygon with area $A_p$.
The sand-free area is simply $A_f = A_p - \omega^2$, i.e.~the remaining
area that is left after subtracting the area of the dune, approximated
as $\omega^2$.

As previously reported [\cite{Duran09,Hersen04b}], we find that the 
spacing between 
dunes takes well-selected values within the same field. 
Additionally, the inter-dune spacing $L$ 
shows no clear trend as a function of the dune size $w$; 
its mean value $\bar L$ is nearly constant over the whole 
width range and only depends on the selected dune field.

This independence between dune size and dune spacing 
is not only a consequence of the uniformity of the spatial distribution of 
dunes but also a special feature of barchan dune fields deeply rooted in the 
dynamics of dune size selection and their spatial distribution. 
For instance, on static dune fields,
such as longitudinal or star dune fields, the inter-dune spacing 
scales with the dune size, i.e.~larger dunes are surrounded by larger 
empty space, 
due to the way sand is redistributed among the dunes. 
In static dune fields, since the annual average of the relative motion 
between dunes is almost zero, they change their size only by their 
influx-outflux balance. 
Therefore, due to mass conservation, a dune accumulates sand and grows only 
by extracting sand from its neighboring dunes that shrink.

In barchan dune fields, dunes are mobile and therefore can collide with each 
other. Next, we present arguments to strengthen the hypothesis that
the interchange of sand due to dune collisions destroys any simple correlation 
between dune size and inter-dune distance and leads to the observed spatial 
uniformity.

\section{Data modeling: simulating the dune field}
\label{sec:DUNE}

Many barchan dune fields arise from the accumulated sand in the sea
shores. For isolated dunes, the sand flux exchanged with the sea shore would
promote their continuous growth. Other mechanisms
at the dune field scale, such as dune collisions, combined with sand exchange
processes between dunes and with their surroundings, enable their stabilization
at the dune field scale [\cite{Duran09}].

To address the problem of dune field stabilization, we start in this
section with numerical simulations of an entire dune field using a continuous 
sand flux balance 
model [\cite{SauermannKroy01,KroySauermann02,SchwaemmleHerrmann05}]. 
This model has already been successfully applied to explain the formation and 
dynamics of isolated barchan 
dunes [\cite{SauermannAndrade03,SchwaemmleHerrmann05,Parteli07a,Parteli07b,Duran10}], 
the formation of transverse dunes [\cite{SchwaemmleHerrmann04}] and 
the transition from barchan to parabolic dunes through vegetation 
growth [\cite{Duran06b}]. A detailed description of the model can be found in 
Refs.~[\cite{SchwaemmleHerrmann04,Parteli07a,Duran10}].

\begin{figure}[htb]
\vspace*{2mm}
\begin{center}
   \includegraphics[width=0.47\textwidth]{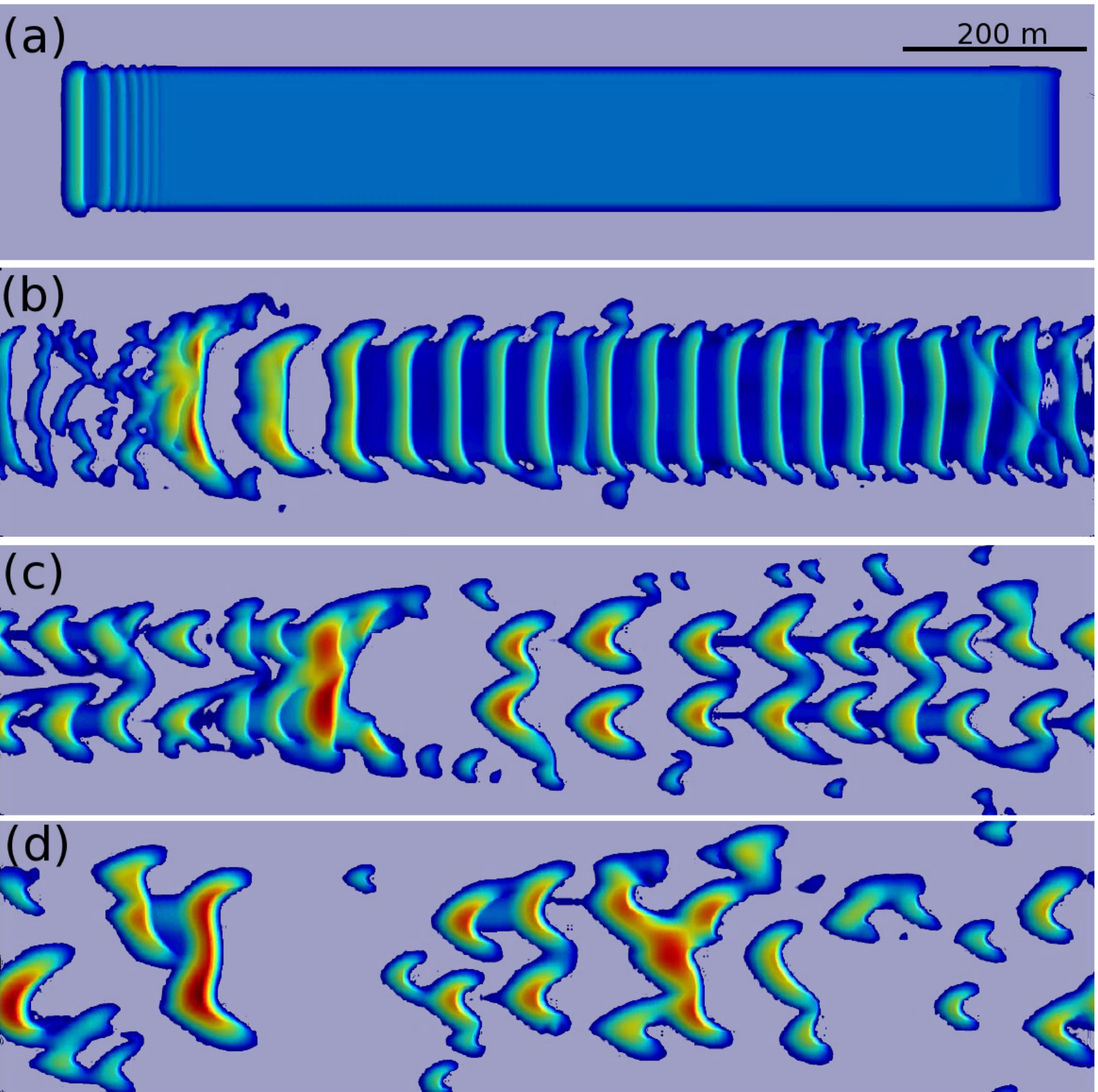}
\end{center}
\caption{(Color online) 
         Snap shots of the dune field evolution beginning with a uniform 
         sand bed under periodic boundary conditions. 
         Wind blows from left to right.} 
\label{fig:fieldDUNESim}
\end{figure}

The model considers a uniform sand bed over a non-erodible surface in the 
center of the field, as illustrated in Fig.~\ref{fig:fieldDUNESim}a. 
Periodic boundary conditions are implemented in both 
the downwind direction and the direction perpendicular to it.

\begin{figure}[htb]
\vspace*{2mm}
\begin{center}
  \fboxrule=0.25mm
  \framebox{\includegraphics[width=0.47\textwidth]{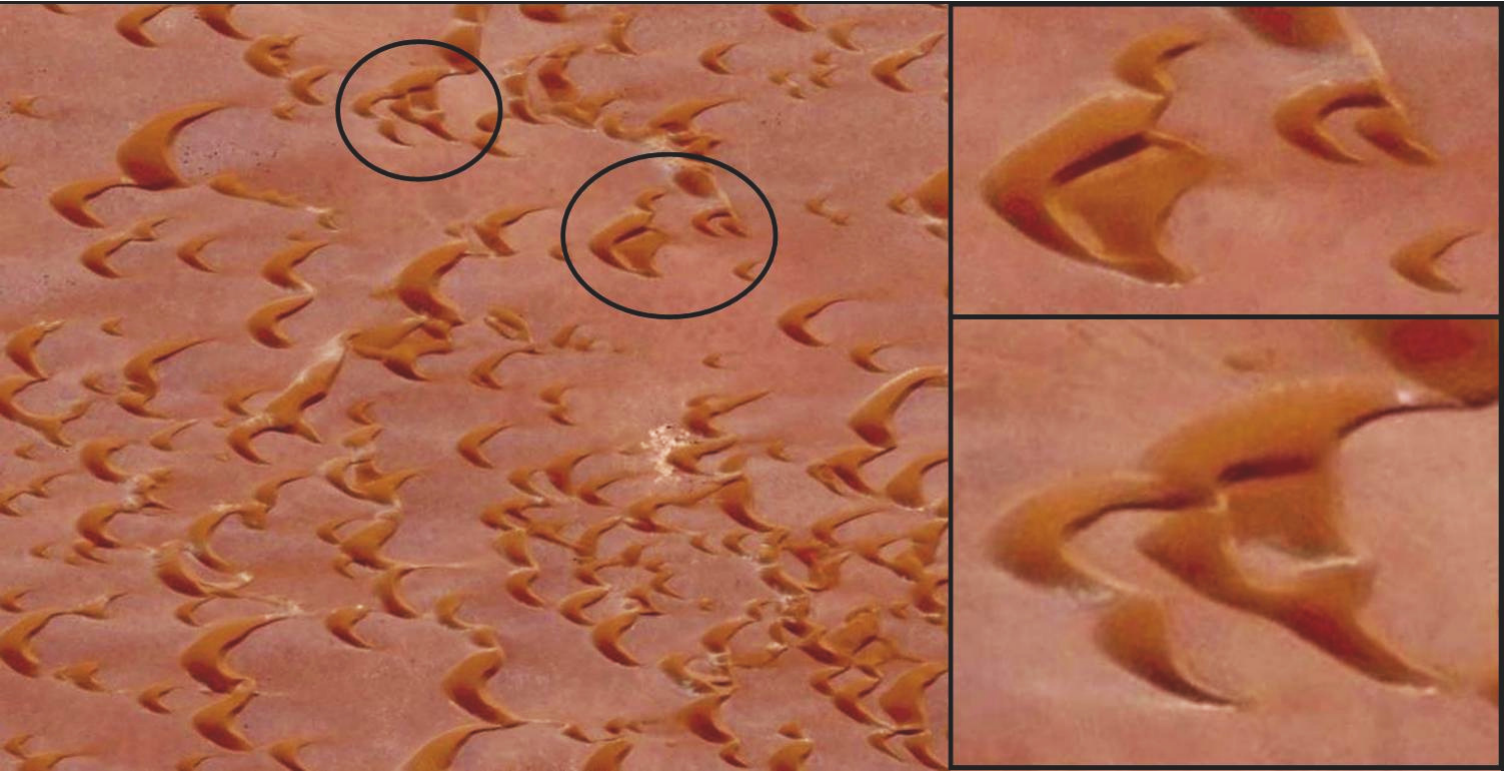}}
\end{center}
\caption{(Color online) 
         Collisions between barchan dunes are ubiquitous in this 
         Moroccan dune field. 
         On the right side are two examples.}
\label{fig:RealCollAll}
\end{figure}

At the beginning, transversal instabilities appear all over the sand bed 
propagating downwind until the whole bed is fragmented into transversal 
dunes (Fig.~\ref{fig:fieldDUNESim}b). Once the sand between the dunes is 
completely eroded, transversal dunes become unstable and split into two 
separated lanes of barchan dunes (Fig.~\ref{fig:fieldDUNESim}c). 
Difference in dune size leads to collisions between barchan dunes that 
together with the flux exchanged between them act as a size selection 
mechanism, leading to a stationary size distribution 
(Fig.~\ref{fig:fieldDUNESim}d). 
This last stage is the one typically observed in real dune fields
(see Figs.~\ref{fig:RealCollAll}),
characterized by the emergence of clusters of colliding dunes 
and alternating localizations of consecutive barchans.

While the simulated field in Fig.~\ref{fig:fieldDUNESim} reproduces the 
main features of a real one, it has typically $\sim 30$ dunes in its 
stationary state, instead of the $\sim 1500$ dunes observed in the real 
fields. Larger number of dunes imply a large computation effort, since
the model reproduces the full shape and dynamics of each dune.

For a proper statistical characterization of size and spatial distributions
in dune fields we propose an alternative model. 
We start in Sec.~\ref{sec:collisions} by describing how collisions between 
dunes may lead to the selection of a characteristic dune size within the
field and in Sec.~\ref{sec:simple} we combine dune collisions
with the full dune motion and the sand-flux exchanged between them.

\subsection{Size regulation by dune collisions}
\label{sec:collisions}

To understand physically the dune size distribution one must take into 
account the dynamical processes that govern the growth of single dunes. 
The intrinsic instability of barchan dunes under an incoming sand flux 
leads to an  increase of the largest dunes in the field whereas the 
smaller ones shrink until they disappear [\cite{Duran10,Hersen04b,Hersen05b}]. 
Hence, the mean size of the dunes should grow with the distance from 
the beginning of a field. Nevertheless, in many dune fields the sizes 
saturate. Two mechanisms have been proposed to avoid unlimited dune 
growth: instability of large dunes due to changing wind 
directions [\cite{Elbelrhiti05}] and collisions between 
dunes [\cite{SchwaemmleHerrmann03,Duran05,Hersen05b}]. 
Here, we concentrate on the second mechanism. 
\begin{figure}[htb]
\vspace*{2mm}
\begin{center}
   \includegraphics[width=0.3\textwidth]{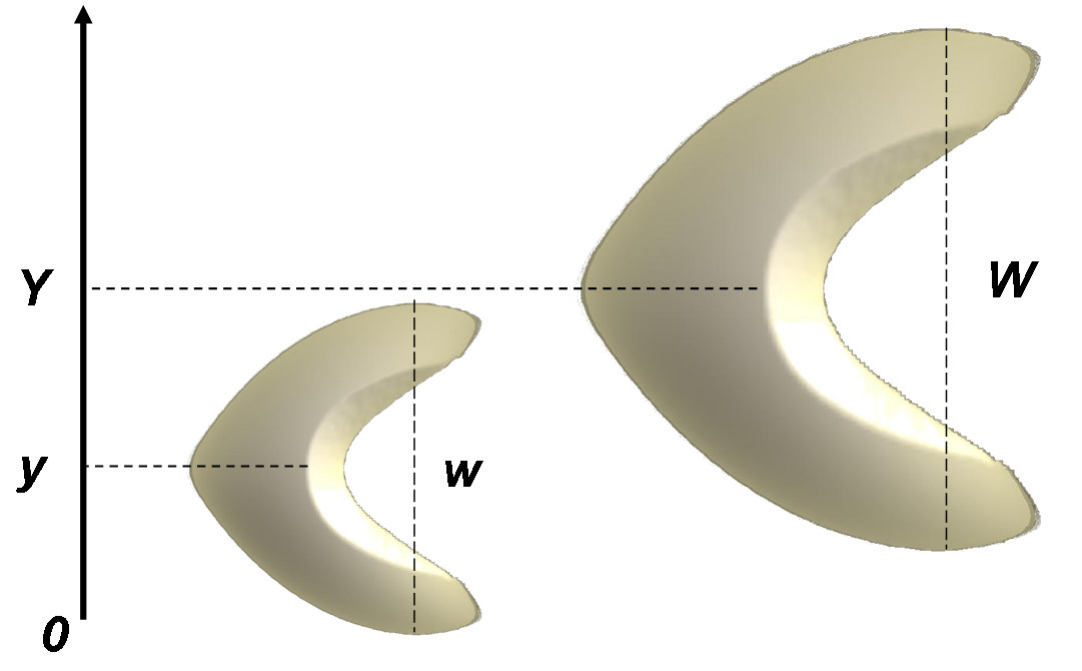}
\end{center}
\caption{(Color online) 
         Sketch of the initial state of a binary collision between two 
         barchan dunes of sizes $w$ and $W$, and centered at $y$ and 
         $Y$, respectively.} 
\label{fig3}
\end{figure}

Collisions are ubiquitous in dune fields (see Fig.~\ref{fig:RealCollAll}) 
due to the relatively broad range of different velocities which obey in 
general $v\sim 1/w$ for single dunes [\cite{Hersen05}]. Due to this 
dependence of their velocity 
on their size, the dune that collides onto a second one must be 
smaller than the latter one. This process has been observed several 
times [\cite{Besler97,Besler02}] but was not understood until recently. 
The large temporal scale of such a process makes it difficult to observe 
the final state after such a collision. 

\begin{figure}[tb]
\vspace*{2mm}
\begin{center}
   \fboxrule=0.25mm
   \framebox{\includegraphics[width=0.47\textwidth]{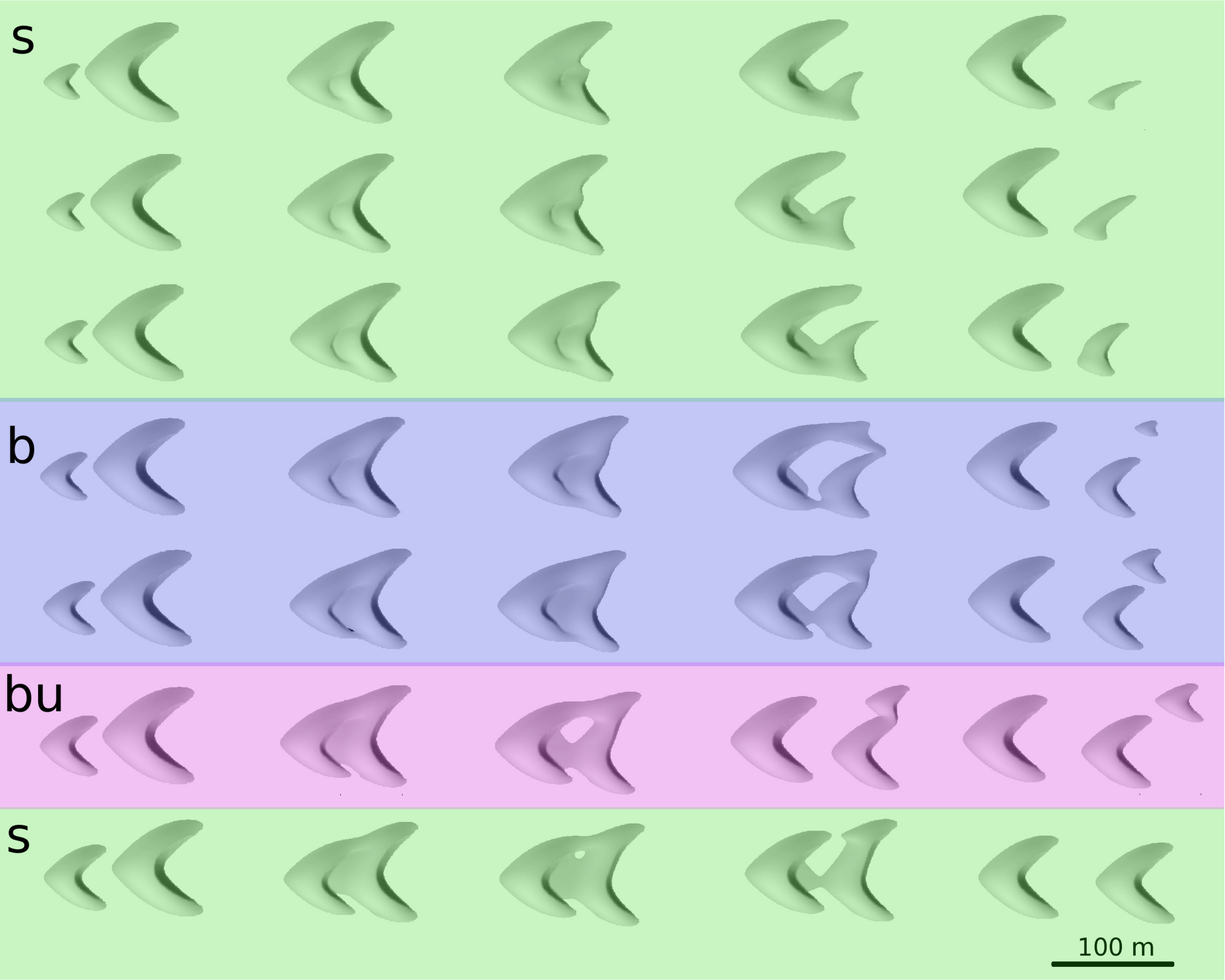}}
\end{center}
\caption{(Color online) 
         Snap shots of the time evolution of binary collision for 
         $\theta_i=0.2$ and volume ratios (from top to bottom): 
         $r_i=0.06$, $0.08$, $0.12$, $0.17$ and $0.3$. Letters and colors 
         distinguish the different results after collision. Notice that 
         the smallest volume ratio $r_i=0.06$ used for the set of 
         simulations, is large enough to avoid dune coalescence.} 
\label{fig4}
\end{figure}

Simulations using the continuous 
dune model were carried out to understand what happens when two dunes collide 
with each other. 
Figure \ref{fig4} shows that, 
after the smaller barchan bumps onto the larger one, 
a hybrid state is formed where the two 
dunes melt into a complex pattern. Depending on the initial relative size 
$r_i \equiv \frac{v}{V}$, where $V$ 
is the volume of the large barchan and $v$ the volume of the small one, 
and their lateral offset $\theta_i \equiv \frac{\vert Y-y\vert}{W/2}$, where $Y$ and $y$ 
are the coordinates of the crest of the large and the small dune in the 
lateral direction transverse to their movement, respectively, and $W$ is 
the width of the larger dune (see Fig.~\ref{fig3}), four different situations can 
emerge after collision: coalescence, where only one dune remains, breeding 
(Fig.~\ref{fig4}, `b') and budding (Fig.~\ref{fig4}, `bu') where two dunes 
leave the larger one, and solitary wave behavior (Fig.~\ref{fig4}, `s') 
where the number of dunes remains two after the collision.
These different final situations provide mechanisms to redistribute sand and thus to 
avoid the continuous growth of dunes in a dune field. Similar occurrences 
can be observed in experiments with sub-aqueous barchans [\cite{Endo04}]. 

\begin{figure}[h]
\vspace*{2mm}
\begin{center}
  \includegraphics[width=0.5\textwidth]{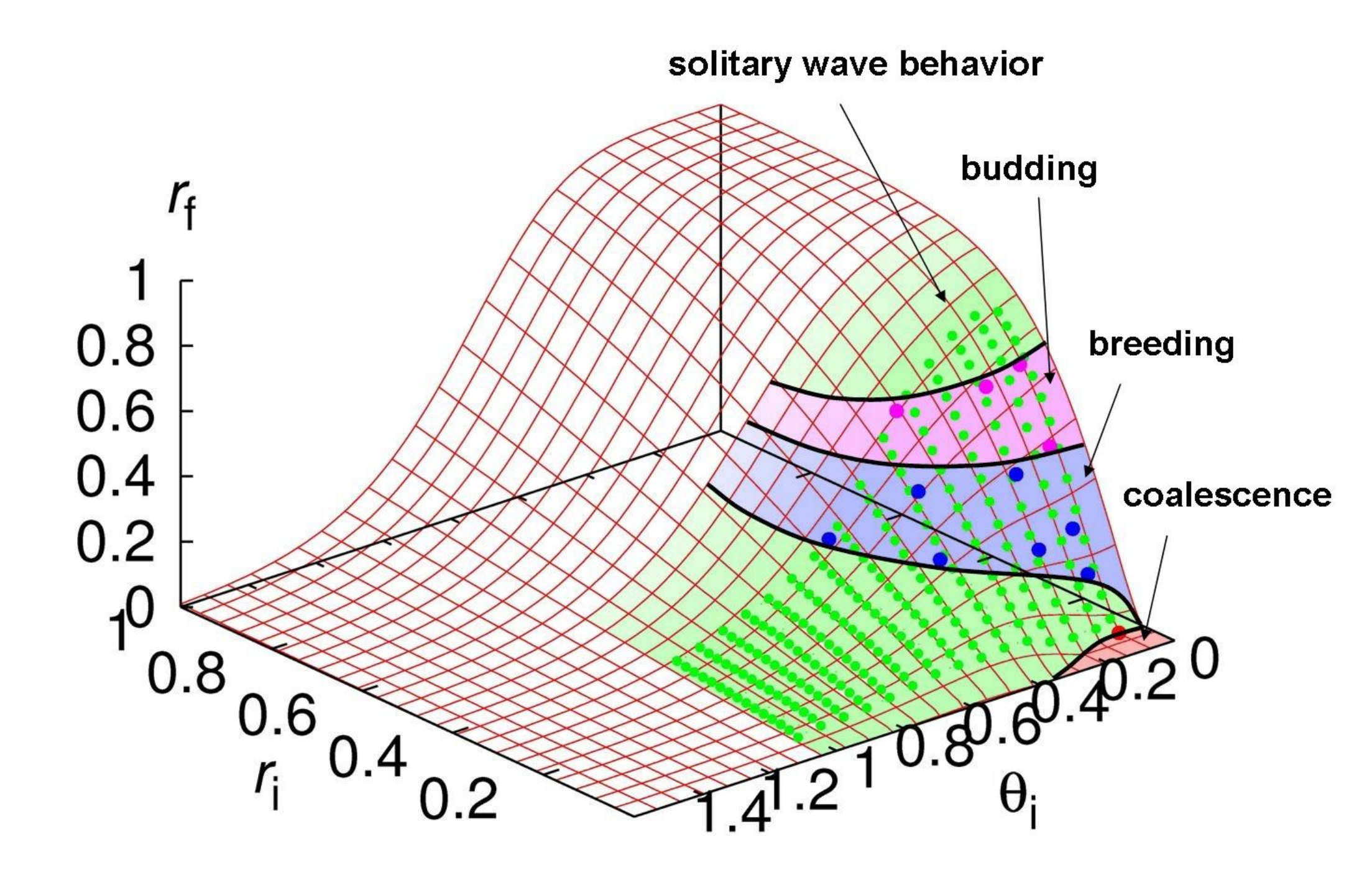}
\end{center}
\caption{(Color online) 
         Sketch of the morphological phase diagram for binary collisions. 
         The volume ratio $r_f$ after the collision is plotted as a 
         function of the initial offset $\theta_i$ and the initial volume 
         ratio $r_i$. Dots represent simulation results.}  
\label{fig:Regimes}
\end{figure}

Next we construct a heuristic set of collision rules based on
simulations with the continuous dune model described above.
These collision rules provide the same statistical output as the
continuous model, enabling to treat pairs of dunes as single objects 
which interact whenever the initial relative lateral offset between them is
$\theta_i < 1$.
Together with the initial lateral offsets we also consider
the corresponding initial volume ratio $r_i$.

The morphological phase diagram of binary collisions is schematically shown 
in Fig.~\ref{fig:Regimes} in terms of the final volume ratio $r_f$ as a 
function of the initial volume ratio and lateral offset. For simplicity only 
conservative collisions are included, i.e. we assume that after the collision 
the summed volume of the ejected dunes corresponds 
to only one characteristic dune.
We notice that the lateral positions of both dunes change after the collision, 
but no simple rule could be found. Therefore, 
a new mutual lateral offset is tossed for each collision.
Appendix \ref{append:collisions} contains all details concerning the
rules for collisions.

Using the diagram in Fig.~\ref{fig:Regimes} for taking the final volume
ratio $r_f$ after one collision, we next consider
the simplest approach to a dune field model, namely, a system that consists 
of a large number of dunes, characterized only by their width, which interact 
exclusively through collisions between them. 
For each collision two dunes are taken randomly from the field to collide. 
This is repeated every iteration as many times as the number of dunes in the 
field. 

Within this framework, we study the evolution of the 
dune size distribution $P_{col}(w)$ in the entire field in order to check if 
the macroscopic behavior of the system approaches a steady state.

We have shown that [\cite{Duran09}] the size distribution function 
converges toward an absorbent state with a stable
Gaussian-like distribution with mean width $\langle w\rangle_{col}$. 
The total mass of all dunes is conserved with the exception of a 
negligible amount due to the small dunes removed from the field. 
Therefore, the mean dune size $\langle w\rangle_{col}$ is determined by 
the average volume $\langle V\rangle$. 

Since it is a Gaussian, the size distribution $P_{col}(w)$ only depends 
on the average volume of the field $\langle V\rangle$, namely
\begin{equation}
\label{rho_c}
P_{col}(w) = \frac{1}{\sqrt{2\pi}\,\sigma_{col}}\exp{
            \left [ -\frac{(w-\langle w\rangle_{col})^2}{2\,\sigma_{col}^2}
            \right ]}.
\end{equation}
Furthermore, the mean square deviation $\sigma_{col}$ is proportional to 
the mean dune size $\langle w\rangle_{col}$ [\cite{Duran09}].

From our findings above one concludes that collisions alone act 
as a random additive process and are able to select a 
characteristic dune size from a given initial condition. 
However, this mean-field approach doesn't give information 
about neither the spatial distribution of the dunes nor the role played by 
the positions of the dunes on the actual collisions.
 
\subsection{A simplified dune field model}
\label{sec:simple}

Calculations of very large dune fields are still difficult because of high 
computational costs. 
The continuous dune model reproduces the dune evolution at the scale of the 
saturation length (typically $\sim 1$m) and thus is extremely expensive in terms 
of running time for large dune field simulations. 
One way out would be to consider a simplified `coarse-grained' dune model,
where dunes are themselves the basic objects.
For that purpose we use the collision rules obtained above
together with the rules for the motion and evolution of barchans 
obtained from continuous simulations [\cite{Duran10}]. 

This effective model considers
a barchan dune field with constant unidirectional wind,
a maximum length $x_{max}$ in the wind direction and width $y_{max}$
and fed by
small barchans entering upwind into the field.
The width $w_0$ of the incoming dunes is constant and they enter 
at a rate $\nu$ --number of incoming dunes per time step. 
Their $y$-position is randomly distributed. 
\begin{figure}[htb]
\vspace*{2mm}
\begin{center}
  \includegraphics[width=0.49\textwidth]{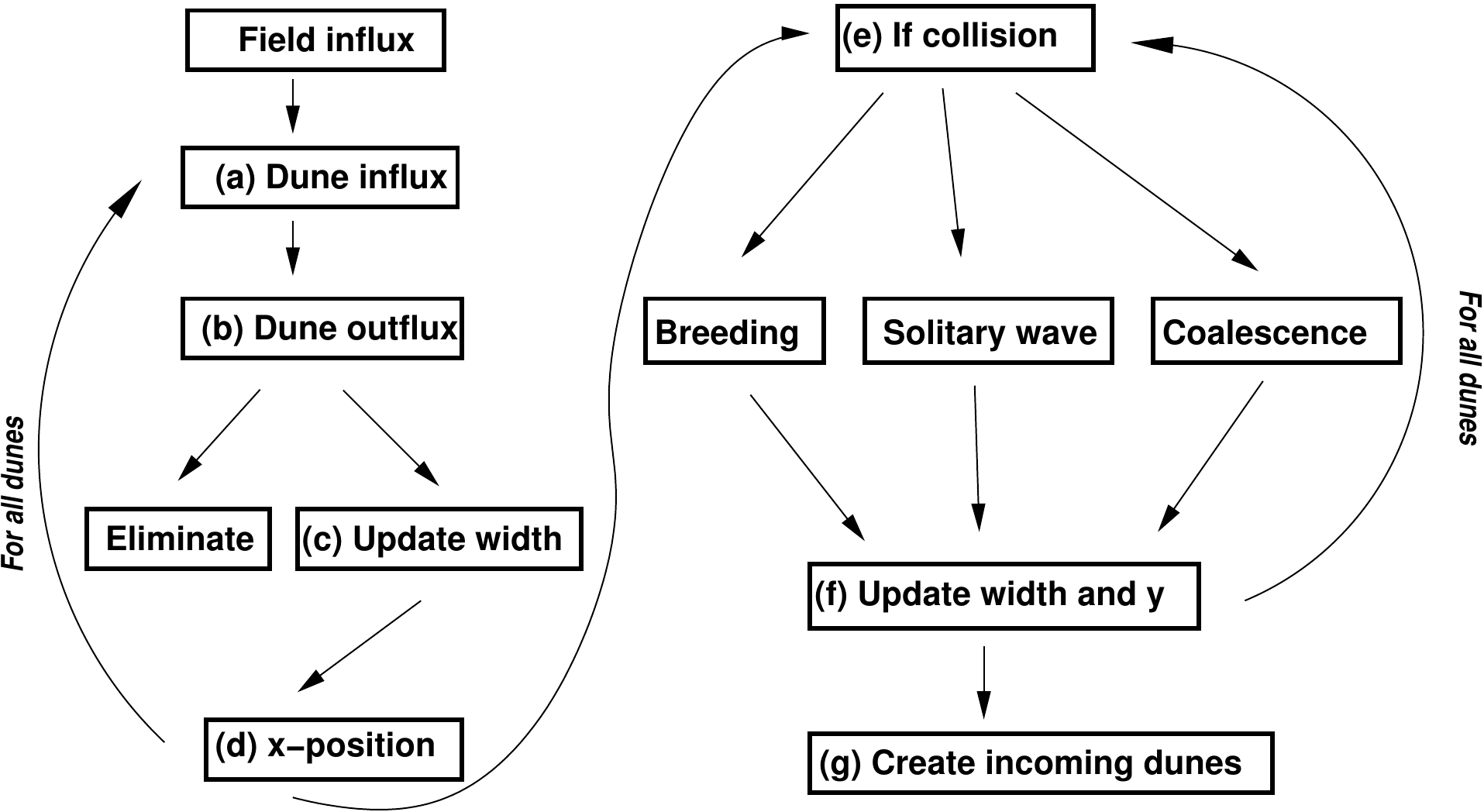}
\end{center}
\caption{The model for barchan dune fields.
         Illustration of the set of operations executed at each time 
         step.}
\label{fig:Illu}
\end{figure}

Each dune is characterized by its width $w$ and its coordinates in the field, 
$x\in [0,x_{max}]$ and $y\in [0,y_{max}]$.
In each iteration the dunes change their size and position due to the sand flux 
balance and collisions. 
Next, a detailed description of the algorithm 
is presented. A graphical illustration of the different 
steps during 
one iteration is given in Fig.~\ref{fig:Illu}.
\begin{table}[htb]
\caption{\label{tab:Tmodel}Model parameters}
\vskip4mm
\centering
\begin{tabular}{@{}ll}
\tophline
{\bf Dune parameters} & \\
Dune outflux-influx relationship, slope: & $a=0.45$\\
Dune outflux-influx relationship, offset: & $b=0.1$\\
Proportionality between volume and cubic width: & $c= 0.017$\\
Dune velocity constant: & $\alpha= 50$\\
\middlehline
{\bf Model parameters} & \\
Time step: & $\Delta t=0.001$ yr \\
Maximum number of iterations: & $T = 10^6\equiv 10^3$ yrs\\
Field width: & $y_{max} = 3$ km\\
Field length: & $x_{max} = 20$ km\\
Saturated flux: & $Q = 300$ m$^2$/yr\\
Dune field influx: & $q_{f,in} = 0$\\
Surface density: & $\rho_0$ (variable) \\
Size of incoming dunes: & $w_0$ (variable)\\
Rate of incoming dunes: & $\nu$ (variable, Eq.~(\ref{rho0}))\\
\bottomhline
\end{tabular}
\end{table}

We start by the sand flux process.
The dune's influx $q_{in}$ determines the volume alteration of a dune and so its new width $w$. Following Ref.~[\cite{Duran10}], 
the mass balance on a barchan dune can be well approximated as 
\begin{equation}
\label{eq:dotw}
\dot w = \frac{Q\delta}{3c\, w},
\end{equation}
where 
$c$ is the proportionality constant between the dune volume and the cubic
power of the width [\cite{Duran10}],
$\delta \equiv (q_{in} - q_{out})/Q$ denotes the flux balance on the dune,
with $q_{in}$ and $q_{out}$ the dune influx and outflux, respectively
and $Q$ is the saturated flux.
Since the normalized outflux can be written as
\begin{equation}
\frac{q_{out}}{Q} = a \frac{q_{in}}{Q} + b ,
\label{normoutflux}
\end{equation}
where $a$ and $b$ are the slope and offset in the outflux-influx 
relation [\cite{Duran10}],  
the flux balance reads $\delta = (1-a)q_{in}/Q-b$.
Table~\ref{tab:Tmodel} indicates the particular values used in our 
simulations.

From time $t$ to $t+\Delta t$ the dune evolves in time with a width 
given by the integration of Eq.~(\ref{eq:dotw}), namely
\begin{eqnarray} 
w(t+\Delta t) &=& \sqrt{w^2(t) + \frac{2 Q\delta}{3 c} \Delta t} 
\label{wtt}
\end{eqnarray}

Meanwhile, the dune moves forward a distance 
$x-x_0$ that results from 
the integration of the dune velocity-width relationship, 
$v=\alpha Q /w$ [\cite{Duran10}].
From Eq.~(\ref{eq:dotw}), this relation becomes
$v=3c\alpha \dot w/\delta$. After integration, it yields
\begin{equation}
\label{eq:dwdx}
w(x) = w(x_0) + \frac{\delta}{3c\alpha} (x - x_0) ~,
\end{equation}
which predicts a linear change of the dune size with the distance 
it moves.

The dune contribution to the sand flux in the field is as follows.
From the normalized outflux in Eq.~(\ref{normoutflux}), the total sand 
flux out of a dune of width $w$ is $q_{out} w$, where the flux is assumed 
to be homogeneously distributed along the dune width, due to diffusion 
processes. 
This is in fact a simplified picture of what happens in real dunes. 
There, the sand leaves the horns with an intensity of nearly the saturated 
flux $Q$ and the remaining part of the dune is dominated by the dune's 
slip face from where almost no sand leaves [\cite{KroySauermann02}].
Thus, on average $q_{out}w$ is a good approximation for the total sand flux.

The updated flux field determines the influx on the next dune. 
This dune again updates the flux field by replacing the influx at the 
corresponding $x$-position by its outflux while simultaneously either 
changing its size or being eliminated from the field.

After updating all dunes according to the actual sand flux of the field at 
their position, we look if their new positions and sizes lead to collisions. 

First, we check if a dune overtakes another one or if they overlap in their 
lateral extension. 
When they overlap, we apply the collision rule, derived in the previous 
section, and calculate the new widths and positions. 
Therefore, collisions are taken as instantaneous and
every time two dunes collide we select a small random lateral offset.

At the end of each iteration, incoming dunes are generated and 
positioned at the beginning of the field, $x=0$.

\section{Results}
\label{sec:results}

\begin{figure*}[htb]
\vspace*{2mm}
\begin{center}
  \includegraphics[width=0.9\textwidth]{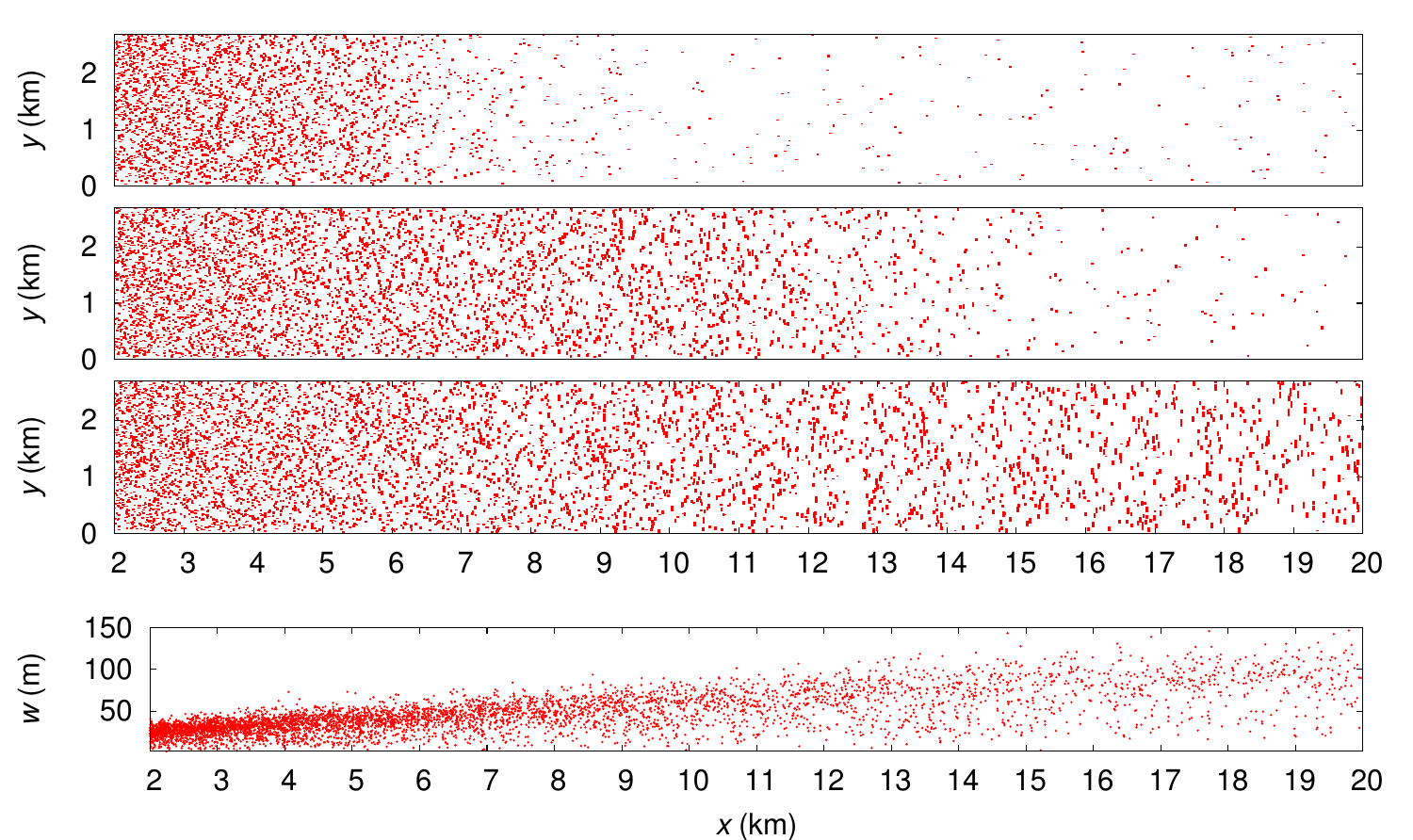}
\end{center}
\caption{
         Top, three characteristic stages of the evolution of a dune 
         field at 20 yr ($2\times 10^4$ steps), 50 yr ($5\times 10^4$ 
         steps) and the steady state after about 100 yr ($10^5$ steps). 
         Dunes move from left to right and are represented by the 
         `width line', i.e. a line centered in the center of mass of 
         the dune and with a length equal to the dune width. Bottom, 
         dune width $w$ as function of its distance downwind at the 
         steady state.}
\label{fig:fieldSim}
\end{figure*}

In this Section we present the main results from simulations
for different dune input rates $\nu$ and sizes $w_0$ using the model,
described in the previous section.

Since the incoming dunes are randomly distributed along the input boundary 
$x=0$ we impose the density $\rho_0$ of the incoming dunes instead of the 
input rate $\nu$. From the definition, $n$ dunes of width $w_0$ uniformly 
distributed in an area $A=y_{max}\Delta X$ have a surface density $\rho_0 
\equiv (n\,w_0^2)/(y_{max}\Delta X)$ where $y_{max}$ is the field width and 
$\Delta X= v\,\Delta t$ is the distance covered by dunes with velocity 
$v=\alpha Q/w_0$ during a time interval $\Delta t$ equal to one time step. 
Since by definition the input rate is 
$\nu=n/\Delta t$, the dune density becomes 
\begin{equation}
\rho_0 = \nu \frac{w_0^3}{\alpha Q y_{max}}
\label{rho0}
\end{equation}
where the parameters are given in table \ref{tab:Tmodel}.
We apply periodic boundary conditions in the direction perpendicular to 
the wind.

Figure ~\ref{fig:fieldSim} (upper panel) shows the evolution of a 
typical dune field with a high density ($\rho_0=0.42$) of about 2 m high 
incoming dunes ($w_0=20$ m). The dune field invades the whole simulated 
area and finally reaches a steady spatial distribution. In general, 
along the wind direction, the spatial distribution is not uniform, 
dunes become progressively sparse, and at the same time the dune size 
increases (Fig.~\ref{fig:fieldSim}, bottom). As will be shown below, 
this coarsening is a direct consequence of the -unstable- flux balance 
and differs from the homogeneous distribution of real dune fields 
(see Fig.~\ref{fig:fieldReal}).

\begin{figure*}[htb]
\vspace*{2mm}
\begin{center}
  \includegraphics[width=0.95\textwidth]{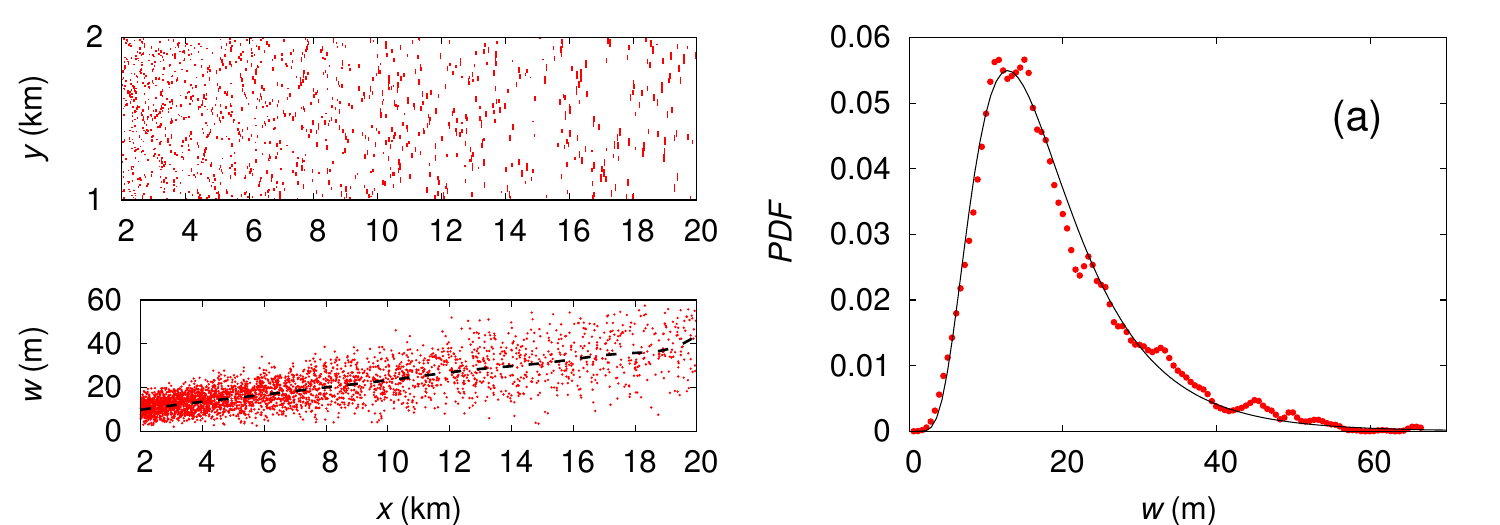}
  \includegraphics[width=0.95\textwidth]{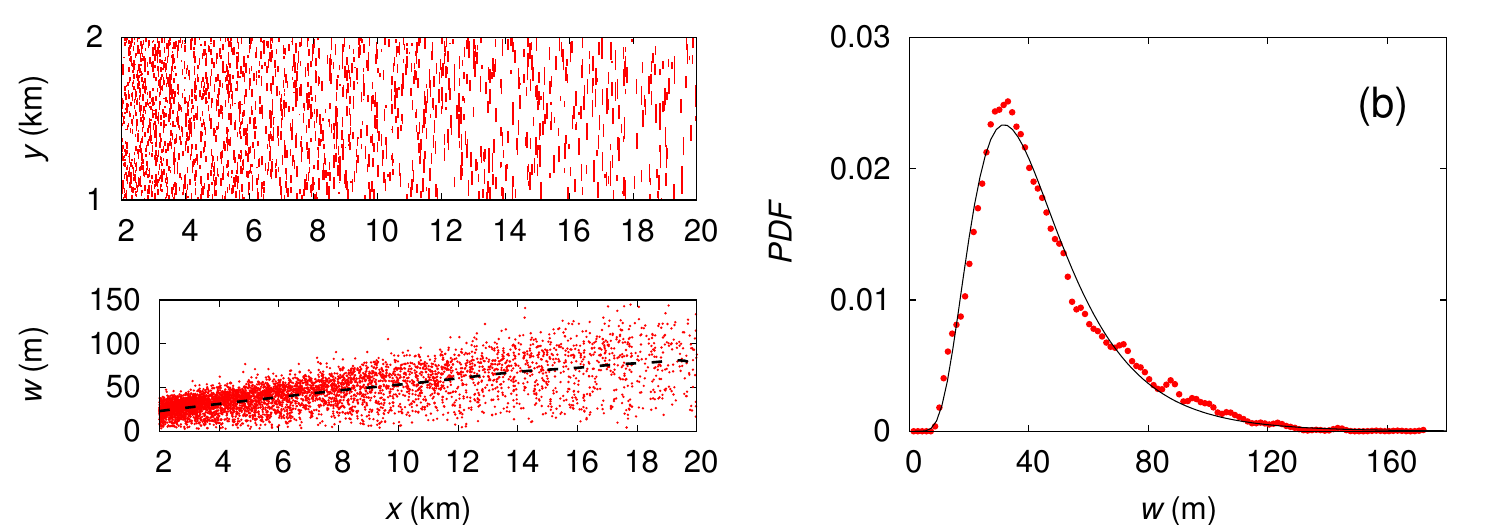}
\end{center}
\caption{Dune width distribution of two simulated dune fields together 
         with a snap shot of the field in the stationary state (top left) 
         and the downwind dune size distribution (bottom left). In the 
         distribution function, dots are measured points and the line 
         is the analytical model. The boundary conditions are: 
         (a) $w_0=10$ m, $\rho_0=0.18$, and (b) $w_0=20$ m, $\rho_0=0.42$.}
\label{fig:fieldSimDist}
\end{figure*}

In spite of this difference, the global dune size distribution in the 
simulated fields actually corresponds to the measured ones, and thus 
are also well described by the analytical mean-field model [\cite{Duran09}]
(Fig.~\ref{fig:fieldSimDist}). Therefore, the Gaussian distribution 
induced by the high rate of collisions at the beginning of the field, 
is gradually skewed toward large sizes due to the coarsening. However, 
the interaction dynamics we use is too simple and does not capture the 
fragmentation process that should compensate coarsening and lead to a 
homogeneous distribution. This shortcoming is discussed in the next 
section, along with some ideas how to overcome it.

\begin{figure}[h!]
\vspace*{2mm}
\begin{center}
  \includegraphics[width=0.5\textwidth]{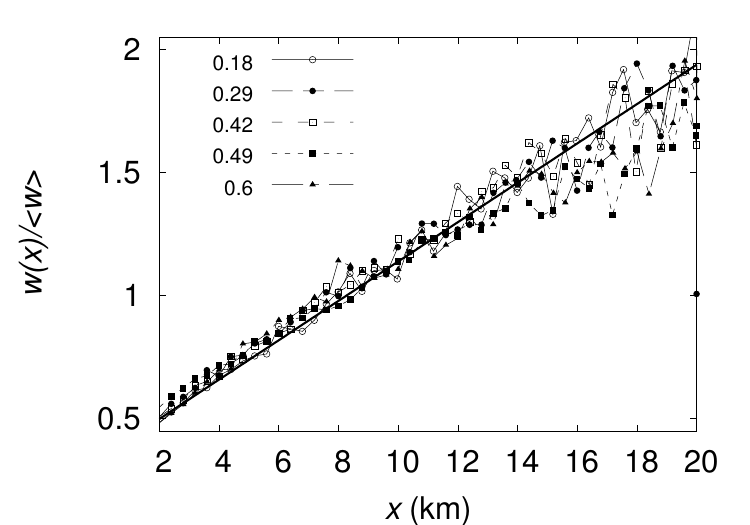}
\end{center}
\caption{Collapse of the normalized local mean dune width  
         $w(x)/\langle w\rangle$ as function of the downwind distance 
         $x$ (symbols) for different boundary conditions, e.g. a 
         different input dune density $\rho_0$. The solid line corresponds 
         to the linear fit. 
         $\langle w\rangle$ is the mean dune size in the whole field.} 
\label{fig:sim_w}
\end{figure}

\begin{figure}[h!]
\vspace*{2mm}
\begin{center}
  \includegraphics[width=0.5\textwidth]{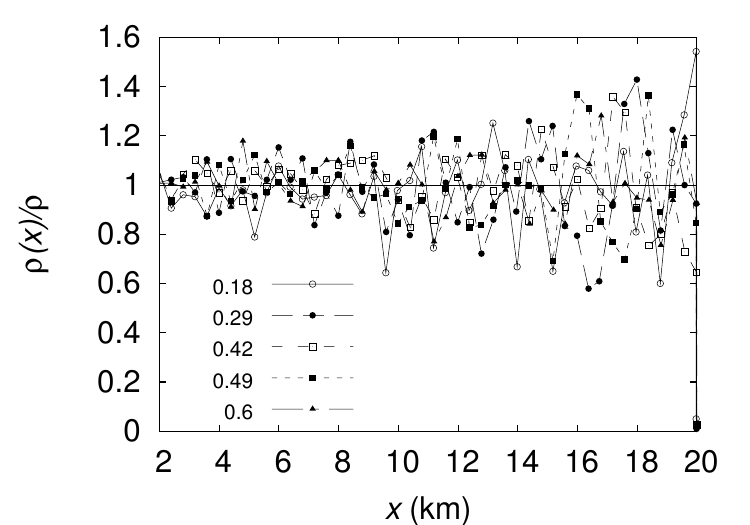}
\end{center}
\caption{The local dune density $\rho(x)$ as function of downwind 
distance remains constant along the whole field independent of the 
boundary conditions. In the figure $\rho(x)$ is normalized by the 
total density $\rho$} 
\label{fig:sim_rho}
\end{figure}

\begin{figure}[htb]
\vspace*{2mm}
\begin{center}
  \includegraphics[width=0.5\textwidth]{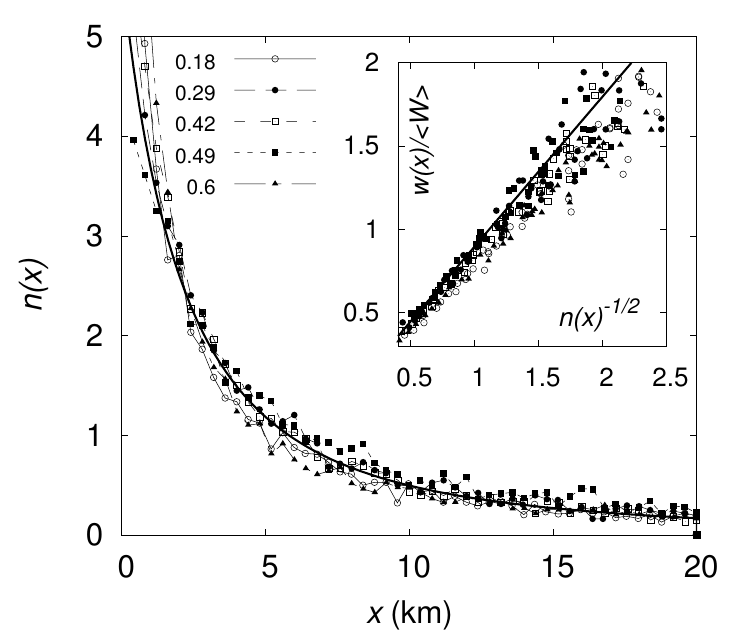}
\end{center}
\caption{Normalized local dune density $n(x)\equiv N(x) (x_{max}/dx)/N$, 
where $N(x)$ is the fraction of the total number of dunes $N$ placed 
between $x-dx$ and $x+dx$, as function of the downwind distance $x$ for 
different input dune densities (symbols). The solid line corresponds to 
the $1/w^2(x)$ fit (see text). Inset: the local dune mean size $w(x)$ 
scales as $\langle w\rangle / \sqrt{n(x)}$ (solid line).} 
\label{fig:sim_n}
\end{figure}

As shown in Fig.~\ref{fig:sim_w} the local average dune size $w(x)$ increases 
linearly with the downwind distance $x$. 
The width $w(x)$ is defined as the average size inside an area 
$[x-dx/2,x+dx/2]\times [0,y_{max}]$, where $dx$ is the length of the averaging 
window. Surprisingly, this increasing average of the dune size is not 
affected by the collision dynamics and simply follows the flux balance 
result given by Eq.~(\ref{eq:dwdx}). 
After normalizing the dune size by the mean width $\langle w\rangle$, all 
curves corresponding to different input dune densities $\rho_0$ collapse. 
This linear increase has also been observed in small real dune fields in 
Morocco and is apparently related to the initial states of the dune 
fields [\cite{Elbelrhiti08}].

Following Eq.~(\ref{eq:dwdx}), from the slope of the spatial increase 
of the mean size $w(x)$ it is possible to calculate the average inter-dune 
balance term $\langle\delta\rangle$. Based on the definition 
$\langle\delta\rangle\equiv (1-a)\langle q_{in}/Q\rangle - b$ and the values 
$a\approx 0.45$ and $b\approx 0.1$, one can then estimate the average 
normalized influx inside a dune field, $\langle q_{in}/Q\rangle$ which is 
in the range $1.04-1.17\,q_c/Q$. Therefore, the average influx is very 
close to the equilibrium influx $q_c/Q$ at which dune outflux equals dune 
influx.
Furthermore, from Fig.~\ref{fig:sim_w} the 
difference $\langle q_{in}/Q\rangle - q_c/Q$ as expected, scales with the 
ratio $\langle w\rangle/L_c$, i.e. a higher mean dune influx implies a 
larger mean size.

Another interesting result that follows from the conservation of sand 
inside the field, is that the local dune density $\rho$ remains constant 
along the field (Fig.~\ref{fig:sim_rho}). The local dune density is 
defined as $\rho(x) = A_s(x)/A$, where $A_s(x)$ is the fraction of the 
local area $A = dx\times y_{max}$ covered by dunes. Since $A_s(x) 
\approx N(x) w^2(x)$, with $N(x)$ as the local dune number with mean size 
$w(x)$, it follows that the local concentration of dunes scales as 
$1/w^2(x)$ (Fig.~\ref{fig:sim_n}).

\begin{figure}[htb]
\vspace*{2mm}
\begin{center}
  \includegraphics[width=0.55\textwidth]{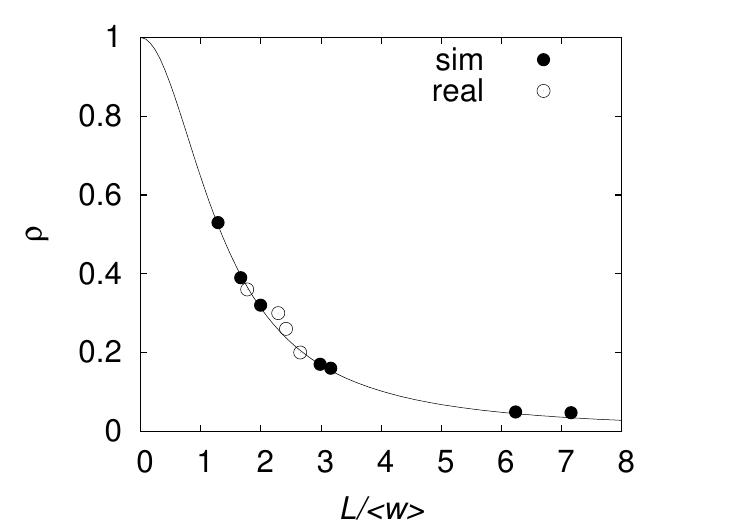}
\end{center}
\caption{Dune density $\rho$ as a function of the relative inter-dune 
spacing $\bar{L}/\langle w\rangle$ for the real dune fields in 
Fig.~\ref{fig:fieldReal} ($\circ$), simulated dune fields ($\bullet$)}
and the fit when both are taken into account (solid lines).
\label{fig:S_rho}
\end{figure}

From the definition of the dune density $\rho \approx N\langle 
w\rangle^2/(N\langle w\rangle^2+A_L)$ with $N$ being the number of dunes 
and $A_L$ being the free total area between dunes, and taking into account 
that $A_L$ scales with the mean inter-dune spacing $\bar L$ as $\bar{L}^2$, 
one may write,
\begin{equation}
\rho = \frac{1}{1+\gamma\left (\frac{\bar{L}}{\langle w\rangle}\right)^2}~.
\label{eq:rho}
\end{equation}
Figure \ref{fig:S_rho} shows the densities of both, measured and simulated 
dune fields, with circles and bullets respectively, as a function of the 
relative inter-dune spacing $\bar L/\langle w\rangle$. The solid line 
indicates a least square fit with the functional form in Eq.~(\ref{eq:rho})
with one single parameter $\gamma$.
All points from empirical and simulated data are well predicted by 
Eq.~(\ref{eq:rho}) whose fit yields a value for the fitting constant 
$\gamma\sim 0.6$. Figure \ref{fig:S_rho} clearly strengthens the 
theoretical derivations above, showing that the density of dunes is 
approximatelly uniform and depends exclusively on the relative
inter-dune spacing.

Making use of our first picture how the field properties relate to each 
other, we finally address the open question of why different dune fields 
may have different densities and characteristic dune sizes. From the 
analysis of the simulated fields we found that the density and average 
width are, in fact, dependent on the boundary conditions, namely the 
input density $\rho_0$ of dunes and the corresponding width $w_0$, 
determining the input of dunes into the field. 

\begin{figure}[tb]
\vspace*{2mm}
\begin{center}
  \includegraphics[width=0.5\textwidth]{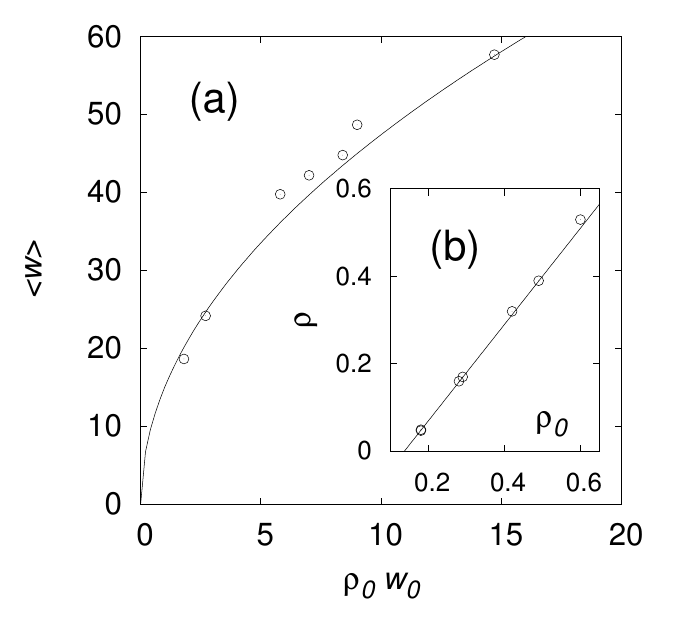}
\end{center}
\caption{\protect
         {\bf (a)} The average width $\langle w\rangle$ as a function
         of the product between the incoming dune density
         $\rho_0$ and the width $w_0$ of the incoming dunes at the
         boundary.
         {\bf (b)} The final stationary density $\rho$ of the dune
         field as function of the initial density $\rho_0$ at the
         boundary. Note that below $\rho_c$ (black circle)
         no dune field formation can be attained.}
\label{fig:Bound}
\end{figure}

For the simulated dune fields, Fig.~\ref{fig:Bound}a suggests the 
following relation for the average width,
\begin{equation}
\langle w\rangle = \left ( \bar{W} \rho_0 w_0 \right )^{1/2} ,
\label{avewidthinit}
\end{equation}
where $\bar{W}\sim 225$ m is a threshold length determining whether the 
mean dune size in the field is smaller or larger than the size of 
incoming dunes. 

In this context, when incoming dunes are smaller than the product 
$\bar{W} \rho_0$, their density is high enough to enhance the sand 
exchange through flux balance and they will grow, increasing the mean 
dune size. Otherwise, if the incoming dunes are larger than $\bar W\rho_0$, 
then their density is too low compared to their size and they cannot 
establish sufficient sand exchange between them. In this case they will 
shrink inside the dune field, decreasing the mean dune size. 

Finally, the dune density $\rho$ in the field, plotted in 
Fig.~\ref{fig:Bound}b, scales with the initial density $\rho_0$ as
\begin{equation}
\rho = \rho_0 - \rho_c~,
\label{rhoinit}
\end{equation}
where $\rho_c\approx 0.12$ is a critical density below which the incoming 
dunes do not receive enough sand to persist and thus disappear. 
Equation (\ref{rhoinit}) can also be understood from volume 
conservation: since the amount of sand blown into the field, given by 
$\rho_0$, has to be shared between the dunes, described by $\rho$, and 
the sand in the aerial layer, one  expects that $\rho<\rho_0$ when the 
corresponding difference $\rho_c$ denotes the density associated to the 
aerial sand flux.   

\conclusions

We presented measurements of dune width and position in four real dune 
fields in Morocco, finding a common underlying size distribution and a 
clear spatial homogeneity. The uniformity of the spatial dune distribution 
gives strong evidence that dune collisions are a non-negligible dynamical 
effect.

In order to reproduce the morphology of dune fields
we revisited two previous models, namely a continuous sand 
flux balance model and a mean-field model. The continuous
model, despite being unable to simulate dune fields as large
as the observed ones, provided accurate statistics for the
different types of binary collisions between dunes and confirmed 
the relevance of dune collisions during the whole field evolution. 
The mean-field model uses the dune collisions output from the continuous 
model to introduce heuristic dune collision rules, from which the
observed dune size distributions are obtained.

We introduced a simplified model for dune 
fields that treats dunes as simple elements described by their width.
This model includes simple collision rules and exchanged flux 
in order to account for the interaction between dunes
and uses basic relations between dune volume, area, outflux and velocity 
in terms 
of dune width. 
As a result, the dune size distribution compares very well with the 
measured ones and converges to a spatially stationary distribution
as observed in real dune fields.

In contrast to measured real fields, the simulated ones are not 
spatially uniform. 
A possible explanation could be that the collision model we use is too 
simple. Indeed, we assume that during collisions, the number of dunes does 
not increase. However, simulations of binary collisions show a rather 
different picture, where quite often a colliding dune is unstable and 
splits into two, a situation called either breeding or budding. 
Such `multiplicative' collisions may counterbalance the coarsening effect 
of the sand flux exchange, thus leading to a uniform steady distribution. 
This is a particularly important point since it has been also observed that 
real barchan collisions may lead to dune fragmentation into several small 
dunes [\cite{Elbelrhiti05}].
Such fragmentation appeals for a different data extraction, since it
implies a width of a potentially asymmetric dune. 
Further, as stated above in Sec.~\ref{sec:results}, if the dune field
is small enough, as it is the case of some Moroccan dune fields,
a linear increase of dune size with downwind distance can be observed.
To address these two points, a comparison to these smaller fields
must be done, which is out of our scope. First, because, as explained in 
Sec.~\ref{sec:measures}, our data
does not allow a resolution fine enough to distinguish the formation
of smaller dunes and dunes in hybrid states such the ones in 
Fig.\ref{fig:RealCollAll}.
Second,
because the linear increase of dune size does not hold when collision
enter into play, which occurs for sufficiently large dune fields, such
as the ones addressed in this study.

Another possible contribution for the non-uniform spatial distribution
at large field distances concerns the time scales taken for collisions.
For both the mean-field approach presented in [\cite{Duran09}] as well 
as our effective model, the collision time-scale is much shorter than 
the characteristic time-scale of the evolving dune field. 
The continuous numerical model for the dune field however presents
a collision time-scale of the same order as the typical time of
dune motion. Considering the fact that collision are not instantaneous
and therefore during one collision both dunes move all together
with a lower velocity - proportional to the inverse of the sum of
their widths - one expects that considering instantaneous collision,
while simplifying an analytical approach, may also contribute for
the non-uniformity of the obtained dune field. Additional investigations
should be made to clarify these points.

We also found that the condition at the dune field input boundary, namely 
the size of incoming dunes and their density, are sufficient to determine 
the main properties of dune fields, the dune density $\rho$, the 
inter-dune spacing $\bar L$ and the mean size $\langle w \rangle$.

It should be emphasized that the input boundary exerts a direct influence 
onto the dune field whereas other quantities like wind strength apparently 
do not have much impact. 
An additional mechanism outside the scope of this manuscript is
dune calving: Strong seasonal winds lead to the instability of large dunes 
in the Moroccan barchan field [\cite{Elbelrhiti05}] and this instability 
leads to dune calving which prevents continuous dune growth.

\appendix

\section{\\ \\ \hspace*{-7mm} Heuristic rules for binary dune collisions}
\label{append:collisions}

In this Appendix we describe in detail the collision rules
discussed above in Sec.~\ref{sec:collisions} from which the plot
in Fig.~\ref{fig:Regimes} is obtained.

We consider two dunes with different sizes, $w_M > w_m$. The largest 
dune is located at $(x_M,y_M)$ and the smallest one at $(x_m,y_m)$. 
The $x$-axis is taken parallel to the wind and thus dune widths align 
parallel the $y$-axis (Fig.~\ref{fig3}).
The initial offset is therefore 
\begin{equation}
\theta_i = 2 \frac{\vert y_M - y_m \vert}{w_M}
\end{equation}
and the initial volume ratio is given by
\begin{equation}
r_i = \left ( \frac{w_m}{w_M} \right )^3 .
\end{equation}

For two dunes to interact, it is necessary that the smallest (fastest)  
dune overtakes the largest one. When this happens the two dunes collide
if their width overlap, namely $y_M+\tfrac{1}{2}w_M > y_m-\tfrac{1}{2}w_m$
and $y_M-\tfrac{1}{2}w_M < y_m+\tfrac{1}{2}w_m$.
It is easy to verify that this latter condition implies $\theta_i<
\theta_c\equiv 1+r_i^{1/3}$.

After a collision, the volume ratio $r_f$ can be approximately expressed by 
the phenomenological equation
\begin{equation}
\label{eq:r_f}
r_f(\theta_i,r_i) \approx
\left[1-e^{-A(\theta_i)\left[r_i-r_0(\theta_i)\right]^{4/3}}\right]~,
\end{equation}
valid for $r_i>r_0(\theta_i)$. 
This condition takes into account that there is a minimal 
relative size $r_0$ of the incoming dune below which no new dune leaves, 
i.e.~coalescence occurs. The coalescence threshold $r_0$ is found to be 
function of the initial lateral offset $\theta_i$, and after fitting the 
numerical data it can be approximated by, 
\begin{equation}
\label{eq:r_0}
r_0(\theta_i) \approx 0.12 \,e^{-(\frac{\theta_i}{0.4})^2} - 0.05~.
\end{equation}
This equation also defines a maximum offset $\theta_i^M\sim 0.4$
above which no coalescence occurs.

On the other hand, the term $A(\theta_i)$ represents the sensibility of 
the final volume ratio $r_f$ to the initial offset and volume ratio, 
and using the numerical data it can be approximated as 
\begin{equation}
\label{eq:A}
A(\theta_i) \approx 10 \left( e^{-(\frac{25\theta_i}{9})^{4/3}} -
                              e^{-(\frac{25\theta_c}{9})^{4/3}}
                       \right ),
\end{equation}

We assume for simplicity that in all collisions there are either one (as for coalescence) 
or two dunes as output (as for solitary wave behavior). The width of the larger one is:
\begin{equation}
\tilde{w}_M = \left (
                \frac{w_M^3+w_m^3}{1+r_f}
              \right )^{(1/3)}
\end{equation}
with $r_f$ given by Eq.~(\ref{eq:r_f}). 
The final offset $\theta_f$ of the outcoming dune is taken randomly 
from the interval $[-1,1]$ (see Sec.~\ref{sec:collisions}).

Coalescence occurs when $r_i<r_0$ (see Eq.~(\ref{eq:r_0}) above).
For larger $r_i$ and $\theta_i$, the slip face survives for longer time, mass 
exchange becomes relevant, and a small barchan is ejected from the main dune. 
We call this process solitary wave behavior. The final width of the smaller dune is
\begin{equation}
\tilde{w}_m=r_f^{1/3}\tilde{w}_M .
\end{equation}
The final offsets of both dunes is
\begin{subequations}
\begin{eqnarray}
\Delta y_M &=& \frac{r_f\theta_f}{1+r_f}-\frac{r_i\theta_i}{1+r_i}
\label{finaloffbigdune}\\
\Delta y_m &=& \frac{\theta_i}{1+r_i}-\frac{\theta_f}{1+r_f}
\label{finaloffsmalldune}
\end{eqnarray}
\end{subequations}
and the corresponding $y$-coordinates are
\begin{subequations}
\begin{eqnarray}
\tilde{y}_M &=& y_M + \sgn{y_M-y_m}\tfrac{1}{2}w_M\Delta y_M ,
\label{ycoordbigdune}\\
\tilde{y}_m &=& y_m + \sgn{y_M-y_m}\tfrac{1}{2}w_m\Delta y_m .
\label{ycoordsmalldune}
\end{eqnarray}
\end{subequations}

\begin{acknowledgements}

PGL thanks {\it Funda\c{c}\~ao para a Ci\^encia e a Tecnologia
-- Ci\^encia 2007} for financial support.
HJH thanks the INCT-SC for support.

\end{acknowledgements}


\addtocounter{figure}{-1}\renewcommand{\thefigure}{\arabic{figure}a}

\end{document}